# Accepted manuscript

# A Comprehensive optimization study of Microbially Induced Carbonate Precipitation (MICP) for soil strength enhancement: impact of biochemical and environmental factors

Yuze Wang [a*], Charalampos Konstantinou [b]

a. Associate professor, Department of Ocean Science and Engineering, Southern University of Science and Technology, Shenzhen, 518055, China；
b. Research associate, Department of Civil and Environmental Engineering, University of Cyprus, Nicosia, 1678, Cyprus

*Corresponding author: Yuze Wang; email address: wangyz@sustech.edu.cn

**Abstract**

Microbially Induced Carbonate Precipitation (MICP) represents a technique for bio-cementation, altering the hydraulic and mechanical properties of porous materials using bacterial and cementation solutions. The efficacy of MICP depends on various bio-chemical and environmental elements, requiring careful considerations to achieve optimal designs for specific purposes. This study evaluates the efficiency of different MICP protocols under varying environmental conditions, employing two bacterial strains: S. pasteurii and S. aquimarina, to optimize soil strength enhancement. In addition, microscale properties of carbonate crystals were also investigated and their effects on soil strength enhancement were analysed. Results demonstrate that among the factors investigated, bacterial strain and concentration of cementation solution significantly influence the biochemical aspect, while temperature predominantly affects the environmental aspect. During the MICP treatment process, the efficiency of chemical conversion through S. pasteurii varied between approximately 80% and 40%, while for S. aquimarina, it was only around 20%. Consequently, the $CaCO_3$ content resulting from MICP treatment using S. pasteurii was significantly higher, ranging between 5% and 7%, compared to that achieved with S. aquimarina, which was about 0.5% to 1.5%. The concentration of the cementation solution also plays a pivotal role, with an optimized value of 0.5 M being critical for achieving maximum efficiency and $CaCO_3$ content. The ideal temperature span for MICP operation falls between 20 and 35°C, with salinity and oxygen levels exerting minor impact. Furthermore, although salinity influences the characteristics of formed carbonate crystals, its effect on Unconfined Compressive Strength (UCS) values of MICP-treated soil remains marginal. Samples subjected to a one-phase treatment, adjusted to pH values between 6.0 and 7.5, exhibit roughly half the UCS strength compared to the two-phase treatment. These findings hold promising potential for MICP applications in both terrestrial and marine environments for strength enhancement.

# 1. Introduction

Microbially Induced Carbonate Precipitation (MICP) is a bio-cementation technique that has become widely adopted in various engineering fields due to its non-disruptive nature. Initially used for soil stabilization (DeJong et al. 2010), liquefaction (Montoya and DeJong 2015), and erosion control (Jiang et al. 2017), it was then expanded in applications relating to water management and fluid flow in porous media applications (Konstantinou et al. 2022; Konstantinou et al. 2023a), for self-healing of soils and cracks (Castro-Alonso et al. 2019), and for groundwater de-contamination. MICP has more recently found applications in marine environments such as erosion resistance under wave actions (Kou et al. 2020), coastal infrastructure construction (Cui et al. 2021) and reinforcement of methane hydrate layers (Hata et al. 2020).

MICP involves a two-step process aiming to strengthen and stiffen porous media such as soil. First, a bacterial solution is applied to the medium, followed by injections of a cementation solution containing urea and a source of calcium. Recently low pH-one phase process was proposed (Cheng et al. 2019; Cui et al. 2022; Lai et al. 2022) in which process bacterial suspension and cementation solutions were mixed and adjusted pH to be lower than 7 to avoid the immediate cementation in the mixed solution. The biochemical process involved in MICP is characterized by two chemical reactions. The first reaction, known as urea hydrolysis, leads to an increase in pH within the system (Equation 1). The positively charged calcium ions attract the negatively charged bacterial walls. The resulting alkaline environment and supersaturation around the bacterial cells cause the precipitation and solidification of calcium carbonate ($CaCO_3$) (Equation 2).

$$CO(NH_2)_2 + 2H_2O \xrightarrow{Urease} 2NH_4^+ + CO_3^{2-} \tag{1}$$

$$Ca^{2+} + CO_3^{2-} \rightarrow CaCO_3(s) \tag{2}$$

This process results in the binding of the particles by the precipitated $CaCO_3$ crystals, thereby enhancing their strength and stiffness. In addition, since the precipitated $CaCO_3$ can be controlled to not fully fill the soil pores, the MICP-treated soil can maintain a relatively high permeability. The properties of bio-treated products are influenced by various factors, including the distribution and morphology of cement within the medium, the characteristics of carbonate crystals (amount, shape, and location), and the base material's particle features (size, shape, roughness, and size distribution) (Konstantinou et al. 2023b). The bio-chemical processes and reactions play a crucial role in determining the rate and characteristics of carbonate precipitation, as there are several mechanisms of crystal growth, transformation, and precipitation that may occur. These mechanisms are affected by several factors, such as bacterial properties (strain, activity, and population) (Omoregie et al. 2017; Wang et al. 2021；Konstantinou et al. 2021b; Tang et al. 2022), the type and concentration of chemicals used in the bio-treatment process, the ratio between chemical amounts (Al Qabany et al., 2012; Al Qabany and Soga 2013; Mahawish et al. 2018; Tang et al. 2022), the interval between injections (Qabany and Soga 2013; Wang et al. 2019, 2021), and the injection method (Martinez et al. 2013; Wu et al. 2019). Therefore, to achieve an optimum MICP design scheme for a particular application, it is necessary to carefully adjust these biochemical factors.

Previously published studies reveal that the concentration of bacteria should be controlled to avoid pore clogging, and sufficient time must be given to allow for bacteria settling prior to cementation injection. To achieve uniform calcium carbonate precipitation, the distribution of bacteria within the medium is critical and depends on both the microbes concentration and the chosen retention period. Martinez et al. (2013) found that unattached bacteria decreased as the distance from the injection source increased regardless of flow direction. Chemical transformation efficiency is maximized by allowing for a retention period between injections, rather than continuous flow. Injection via gravity produces more uniform results than other methods, since it allows for self-adjustment of the flow path. The loading rate in terms of chemical concentrations determines the frequency and time between two subsequent injections. Chemical loading rates less than 0.042 mole/L/hr contribute to high chemical efficiencies (Al Qabany et al. 2012). Due to the phase transformation of $CaCO_3$, higher injection rate promotes the precipitation of more stable and larger $CaCO_3$ crystals which are more efficient in binding soil particles (Wang et al. 2019, 2021). The choice of bacteria with lower urease activity promotes more uniform precipitation by balancing slower MICP reactions (Konstantinou et al. 2021b).

The performance of MICP can be influenced by various environmental factors, such as the pH level, temperature, oxygen levels, and salinity of the aqueous environment (Mortensen et al. 2011; Soon et al. 2014; Kim et al. 2018; Li et al. 2018; Peng and Liu 2019; Tang et al. 2022). These factors are crucial to consider when applying MICP to deep soils or marine soils. Different bacterial strains have been evaluated for MICP at different temperature ranges, with some strains being temperature-independent while others produce crystals with distinct characteristics (Sun et al. 2019; Hata et al. 2020; Wang et al. 2023). The oxygen levels also affect the performance of MICP, and stimulating aerobic bacteria typically result in greater strength enhancement than stimulating anaerobic bacteria (Pakbaz et al. 2022). According to Kim et al. (2018), a pH value of 7.0 leads to more carbonate crystals being precipitated. In saline environments, high-salinity-tolerant, urease-active bacteria need to be flushed to precipitate insoluble and semi-soluble carbonate salts (Cheng et al. 2014).

To create an effective MICP protocol, a combination of factors must be considered, including bio-chemical factors, external/environmental factors and MICP protocol factors. These factors include injection rates and methods of injection, chemical concentrations, time between consecutive injections, and bacterial properties such as strain, population, and activity, all of which should be selected based on the external environmental conditions such as temperature, oxygen levels, and salinity. However, previous studies tend to focus on only one parameter or a few at a time, while keeping the remaining factors constant in order to isolate the effects of the parameter being examined. Additionally, the parameters chosen for each MICP protocol have varied across different studies, making it difficult to combine the findings, evaluate the interactions between parameters, and create an optimum MICP formulation based solely on the literature.

The primary goal of this paper is to investigate the impact of various dominant parameters on MICP performance. This investigation includes assessing the chemical transformation efficiency, flow rates, uniformity and characteristics of carbonate crystals distribution, and resulting strength, in order to analyze the interaction levels between different bio-chemical and environmental parameters. Specifically, this study presents a comprehensive MICP program that varies several parameters one at a time, including temperature, salinity levels, oxygen conditions, pH levels, bacterial strains and densities, cementation solution concentrations, retention times, and injection methods (one-phase/two-phase). Since such a wide range of parameters is examined under a single MICP protocol, the overall significance of each parameter on the MICP performance is evaluated. By identifying the different effects of these parameters, the study enables the design of optimum MICP procedure for effective soil strength enhancement.

## 2. Materials and methods

**Bacterial strains and cultivation**

The bacterium strain *S. pasteurii* is commonly used in ureolysis-based MICP due to its relatively high urea hydrolytic activity. The *S. pasteurii* strain utilized in this study was purchased from CGMCC (CGMCC1.3687). *S. aquimarina* is another urea-hydrolyzing bacterium derived from the marine environment, and there have been a few studies on its use in MICP (Hata et al. 2013, 2020). The *S. aquimarina* strain (CGMCC 1.3644) utilized in this study was also purchased from CGMCC. Both bacterial strains were prepared using freeze-dried stocks following the method presented by Wang et al. (2019b). After bacterial defrost, activation, and agar medium cultivation, liquid medium cultivation was conducted at 30°C at a shaking rate of 200 rotations per minute (rpm) to achieve bacterial strains with an optical density measured at a wavelength of 600 nm ($OD_{600}$) of 3.0 for both bacteria. For *S. pasteurii*, the cultivation liquid medium used was ATCC 1376 $NH_4$-YE liquid medium, containing 20 g/L yeast extract, 10 g/L ammonium sulfate, and 0.13 M Tris base. For *S. aquimarina*, tryptone-soytone medium was utilized, which contained 15.0 g/L tryptone, 5.0 g/L soytone, and 5.0 g/L sodium chloride. Three different bacterial densities were used, with $OD_{600}$ values of 1.0, 2.0, and 3.0. The bacterial suspension with a lower $OD_{600}$ was obtained by diluting the bacterial suspension using their liquid cultivation medium. The ureolysis activities of bacteria were tested using the conductivity measurement following the procedure given by Whiffin (2007). The tested bacterial activity for *S. pasteurii* at an $OD_{600}$ of 1.0 was approximately 39.4±0.5 mM/h, while the activity of *S. aquimarina* under the same condition was approximately 8.9± 0.5 mM/h.

**Chemical components**

To create the cementation solution for MICP treatment, a mixture of calcium chloride dihydrate, urea, and nutrient broth was dissolved in deionized water. Four different concentrations of calcium chloride were used, namely 0.25 M, 0.5M, 1.0 M, and 1.5 M. The concentration of urea used was 1.5 times higher than that of calcium chloride. The concentration of nutrient broth was kept constant at 3 g/L. The effects of seawater salinity on MICP were studied using artificial seawater, which contains 24.53 g/L of NaCl, 0.695 g/L of KCl, 4.09 g/L of $NaSO_4$, 1.16 g/L of $CaCl_2$, 5.2 g/L of $MgCl_2$, 0.101 g/L of KBr, 0.201 g/L of $NaHCO_3$ and 0.027 g/L of $H_3BO_4$. The artificial seawater was mixed with the cementation solution to achieve the desired final concentration. The seawater-based bio-cemented products were compared against the specimens generated with a cementation solution in which DI water was used and acted as the control experiments. The effects of pH on MICP were studied using HCl and NaOH to adjust the pH of the mixed bacterial and cementation solutions to 6, 7.5, and 9. All chemicals used in the study were of analytical reagent grade.

**Sand properties and specimen preparation**

In this study, CHINA ISO standard silica sand was utilized, and its key characteristics and particle size distribution can be obtained from Wang et al. (2023). The sand had a mean grain diameter of 0.125 mm and a coefficient of uniformity ($C_u$) of 1.4. Based on the Unified Soil Classification System (ASTM, 2017), the sand was classified as poorly graded. To create the specimens, a split acrylic cylindrical mold with a height of 80 mm and an inner diameter of 38 mm was used. The dry sand weight was determined based on a targeted relative density (RD) of 50%, and it was poured into the columns in three stages using the dry pluviation technique.

**MICP treatment procedures**

This study employed both one-phase and two-phase injection methods to examine the effects of bacterial, chemical, injection, and environmental factors on soil strength enhancement via MICP. Bacterial suspension and cementation solution were injected into the soil column using gravity filtration, following the protocols by Al Qabany and Soga (2012) and Wang et al. (2022). For the two-phase injections, bacterial suspension was first injected, and after a settling time was given, cementation solution was injected. In one-phase injection, bacterial suspension was mixed with cementation solution first, and then the pH of the mixed solution was adjusted before the injection. Before the MICP treatment, soil samples were saturated with DI water using the methods presented by Wang et al. (2022). The effects of temperature and oxygen conditions on MICP were also investigated. The temperature was regulated using a temperature-controlled water bath, and anaerobic conditions were achieved by placing the soil samples in an anaerobic chamber. The experimental conditions are summarized in Tables 1, 2 and 3 (bio-chemical factors, environmental factors and injection methods, respectively). In total, 25 groups of tests were prepared for the experiments. In this study, the flow rate of bacterial suspension and cementation solution was measured by determining the time required to inject one pore volume of the soil column. After the MICP treatments were completed, excess soluble salts were flushed out from the soil samples using two pore volumes of deionized water (Whiffin et al. 2007; Dejong et al.2010; Wang et al. 2023). Flushing water after MICP treatment completion has been shown to remove excess nutrients in simulated seawater-based bio-cementation as well (Cheng et al. 2014). The specimens were then removed from the columns and oven-dried at 105 ℃ for at least 24 hours (ASTM 2014; Wang et al. 2023) prior to conducting UCS tests, calcium carbonate content measurement and SEM imaging.

Table 1 Bio-chemical factors

| Test No. | Bacterial strain | BS density $OD_{600}$ | Bacterial settling time (h) | CS concentration (M) | Temperature (°C) | pH | Salinity | Anaerobic/ aerobic conditions | Bacteria injection number |
|---|---|---|---|---|---|---|---|---|---|
| $OD_{600}$ | | | | | | | | | |
| 1 | S. pasteurii | 1 | 24 | 0.5 | 20 | NA | No | Aerobic | 1 |
| 2 | S. pasteurii | 2 | 24 | 0.5 | 20 | NA | No | Aerobic | 1 |
| 3 | S. pasteurii | 3 | 24 | 0.5 | 20 | NA | No | Aerobic | 1 |
| Bacterial retention time | | | | | | | | | |
| 4 | S. pasteurii | 1 | 2 | 0.5 | 20 | NA | No | Aerobic | 1 |
| 5 | S. pasteurii | 1 | 6 | 0.5 | 20 | NA | No | Aerobic | 1 |
| 6 | S. pasteurii | 1 | 12 | 0.5 | 20 | NA | No | Aerobic | 1 |
| 1 | S. pasteurii | 1 | 24 | 0.5 | 20 | NA | No | Aerobic | 1 |
| Bacterial strain | | | | | | | | | |
| 1 | S. pasteurii | 1 | 24 | 0.5 | 20 | NA | No | Aerobic | 1 |
| 7 | S. aquimarina | 1 | 24 | 0.5 | 20 | NA | No | Aerobic | 1 |
| Concentration of cementation | | | | | | | | | |
| 8 | S. pasteurii | 1 | 24 | 0.25 | 20 | NA | No | Aerobic | 1 |
| 1 | S. pasteurii | 1 | 24 | 0.5 | 20 | NA | No | Aerobic | 1 |
| 9 | S. pasteurii | 1 | 24 | 1 | 20 | NA | No | Aerobic | 1 |
| 10 | S. pasteurii | 1 | 24 | 1.5 | 20 | NA | No | Aerobic | 1 |

Table 2 Environmental factors

| Test No. | Bacterial strain | BS density $OD_{600}$ | Bacterial settling time (h) | CS concentration (M) | Temperature (°C) | pH | Salinity | Anaerobic/ aerobic conditions | Bacteria Injection number |
|---|---|---|---|---|---|---|---|---|---|
| Salinity-bacteria | | | | | | | | | |
| 1 | S. pasteurii | 1 | 24 | 0.5 | 20 | NA | No | Aerobic | 1 |
| 11 | S. pasteurii | 1 | 24 | 0.5 | 20 | NA | Yes | Aerobic | 1 |
| 12 | S. aquimarina | 1 | 24 | 0.5 | 20 | NA | No | Aerobic | 1 |
| 13 | S. aquimarina | 1 | 24 | 0.5 | 20 | NA | Yes | Aerobic | 1 |
| Temperature, salinity and oxygen | | | | | | | | | |
| 14 | S. pasteurii | 1 | 24 | 0.5 | 4 | NA | No | Aerobic | 1 |
| 15 | S. pasteurii | 1 | 24 | 0.5 | 10 | NA | No | Aerobic | 1 |
| 17 | S. pasteurii | 1 | 24 | 0.5 | 10 | NA | Yes | Aerobic | 1 |
| 1 | S. pasteurii | 1 | 24 | 0.5 | 20 | NA | No | Aerobic | 1 |
| 18 | S. pasteurii | 1 | 24 | 0.5 | 20 | NA | No | Anaerobic | 1 |
| 19 | S. pasteurii | 1 | 24 | 0.5 | 20 | NA | Yes | Anaerobic | 1 |

Table 3 MICP protocols considering bacterial injection number and one-phase injection

| Test No. | Bacterial strain | BS density $OD_{600}$ | Bacterial settling time (h) | CS concentration (M) | Temperature (°C) | pH | Salinity | Anaerobic/ aerobic conditions | Bacteria injection number |
|---|---|---|---|---|---|---|---|---|---|
| Bacterial injection number at low temperature | | | | | | | | | |
| 1 | S. pasteurii | 1 | 24 | 0.5 | 4 | NA | No | Aerobic | 1 |
| 20 | S. pasteurii | 1 | 24 | 0.5 | 4 | NA | No | Aerobic | 3 |
| 21 | S. pasteurii | 1 | 24 | 0.5 | 4 | NA | No | Aerobic | 6 |
| Bacterial injection number for *S. aquimarina* | | | | | | | | | |
| 7 | S. aquimarina | 1 | 24 | 0.5 | 20 | NA | No | Aerobic | 1 |
| 22 | S. aquimarina | 1 | 24 | 0.5 | 20 | NA | No | Aerobic | 3 |
| 23 | S. aquimarina | 1 | 24 | 0.5 | 20 | NA | No | Aerobic | 6 |
| One-phase injection | | | | | | | | | |
| 24 | S. pasteurii | 1 | 24 | 0.5 | 20 | 6 | No | Aerobic | NA |
| 25 | S. pasteurii | 1 | 24 | 0.5 | 20 | 9 | No | Aerobic | NA |
| 26 | S. pasteurii | 1 | 24 | 0.5 | 20 | 7.5 | No | Aerobic | NA |

**Calcium concentration measurement**

Calcium concentration of the outflow during the injections of cementation solution was measured using the EDTA titration method. To conduct the titration, first, use a pipette to transfer 50 ml of the sample to a 250 ml Erlenmeyer flask. Then, 2 mL of 2 mol/L sodium hydroxide solution and approximately 0.2 g of dry calcium carboxylic acid indicator powder were added. After mixing the solution, titration began immediately by adding disodium EDTA solution from a burette while shaking the flask constantly. Initially, the titration speed should be slightly faster and gradually slow down as the end point is approached. It is recommended to pause for 2-3 seconds between each drop and ensure full mixing until the solution changes from purple to bright blue, indicating that the end point has been reached. The entire titration process was completed within 5 minutes. The volume of disodium EDTA solution consumed in milliliters was recorded. The calcium concentration c (mg/L) was calculated using Equation 3.

$$c = \frac{c_1 v_1}{v_0} \times A \tag{3}$$

where, c is the concentration of disodium EDTA solution, mmol/L; $v_1$ is the volume of disodium EDTA solution consumed during titration, ml; $v_0$ is the volume of sample, ml; A is atomic mass of calcium (40.08).

**Unconfined compressive strength (UCS)**

Unconfined compressive strength (UCS) tests are commonly used in studies involving MICP because the treated soil samples resemble rock-like specimens that are stronger than conventional soils. To compare with previous research findings, this study also conducted UCS tests. To ensure accurate UCS results, the top and bottom parts of the samples were trimmed to remove any potentially disturbed or uneven zones. The UCS experiments followed the standard test method for intact rock core specimens outlined in ASTM D2938-86 and ASTM D7012-14e1. An axial load was applied at a constant rate of 1.14 mm/min. The height-to-diameter ratios were approximately 2:1, and any deviations were corrected using Equation (4) as suggested by the ASTM D2938-86 standard test method. It should be noted that moisture conditions can affect the UCS results of tested samples, hence the practice of oven-drying the MICP-treated soil samples.

$$UCS = \frac{UCS_m}{0.88 + \left(0.24 \frac{D}{H}\right)} \tag{4}$$

where UCS represents the computed compressive strength an equivalent H/D = 2 specimen, $UCS_m$ represents the measured compressive strength, D represents the core diameter, and H represents the height of the specimen.

**Calcium carbonate content (CCC) measurement**

To determine the $CaCO_3$ content (CCC) of bio-cemented sand, the ASTM Method (ASTM 2014) was utilized. Initially, a calibration curve was established by introducing various quantities of calcite ($CaCO_3$) with hydrochloric acid and measuring the $CO_2$ pressure. A linear relationship was obtained between $CO_2$ pressure and $CaCO_3$ content (Equation 4). Following the completion of UCS tests, the specimens were carefully removed from the testing machine and 15 to 25 g of samples were collected for every 15 mm along the sand column height grounded and placed in the $CaCO_3$ measurement chamber. A container containing 30 ml of 3 M hydrochloric acid was also placed in the chamber without touching the soil specimens. The $CaCO_3$ measurement chamber was completely sealed and gently shaken to aid the reaction between $CaCO_3$ and hydrochloric acid. $CO_2$ produced from the reaction between $CaCO_3$ and hydrochloric acid increased the chamber pressure, which was measured using a pressure gauge inserted at the top of the chamber. The reading was recorded when the pressure value indicated by the gauge no longer changed, and the quantity of $CaCO_3$ was determined using Equation (5).

$$CaCO_3(g) = 0.034 \cdot CO_2 \text{ pressure} + 0.0198 \quad (5)$$

**Chemical Conversion Efficiency**

The chemical conversion efficiency of MICP was determined by dividing the mass of $CaCO_3$ that was precipitated in the sand by the calculated mass of $CaCO_3$ from the cementation solutions, as defined in previous studies (Al Qabany et al. 2012; Wang 2018):

$$Efficiency(\%) = \frac{m(CaCO_3)/m_1(\text{sand})}{c(CaCl_2) \cdot V \cdot M(CaCO_3)/m_2(\text{sand})} \times 100\% \quad (6)$$

where, $m(CaCO_3)/m_1(\text{sand})$ represents the $CaCO_3$ content measured, $c(CaCl_2)$ is the concentration of $CaCl_2$ in the cementation solutions, $V$ is the total volume of cementation solution injected into the samples, $M(CaCO_3)$ is the molar mass of $CaCO_3$ (100 g/mol), and $m_2$ (sand) is the dry mass of sand used to prepare sample columns. It is important to note that the calculated chemical conversion efficiency may underestimate the actual value because Equation (6) does not account for the $CaCO_3$ precipitated on the surface of the mold and in the filter layers located at the top and bottom of the sample.

**SEM images**

After UCS tests, small samples taken from the middles of UCS samples and were prepared for scanning electron microscopy (SEM) imaging using a PHENOM XL (Thermo Scientific) scanning electron microscope to observe the microscale properties of the CaCO3 crystals formed after the MICP treatment.

# 3. Results

The presentation of the results is structured as follows. First, the outcomes of the tests focusing on biochemical factors are presented. This encompasses bacterial density, intervals between consecutive injections, bacterial strain variations, and concentrations of the cementation solution. Subsequently, results of the effects of environmental factors, including salinity, temperature, pH, and oxygen levels, are detailed. The interplay between these environmental factors and the various biochemical elements is then discussed, leading to the development of various MICP protocols aimed at optimizing the characteristics of the generated specimens.

**Bacterial density, retention time and bacterial strains**

*S. pasteurii* is the predominant bacterial strain utilized for MICP, and several bacterial densities have been investigated. This research selected $OD_{600}$ values of 1.0, 2.0, and 3.0 to compare their influence on chemical transformation efficiency, flow rate, calcium carbonate content, and soil strength following MICP treatment. Increasing the $OD_{600}$ from 1.0 to 2.0 and to 3.0, increases the overall calcium conversion efficiency and consequently increases the overall $CaCO_3$ content and the strength of MICP-treated soils (Figure 1). Additionally, the morphology of $CaCO_3$ crystals is influenced by the optical density. SEM images showed that changes in the $OD_{600}$ from 1.0 or 2.0 to 3.0 altered the crystal morphology. At $OD_{600}$ values of 1.0 or 2.0, the predominant $CaCO_3$ crystal type produced was rhombohedral, with these crystals tending to cluster together, whereas when the $OD_{600}$ was 3.0, two primary kinds of $CaCO_3$ crystals were generated: large, round, rough crystals and smaller, elongated crystals (Figure 2).

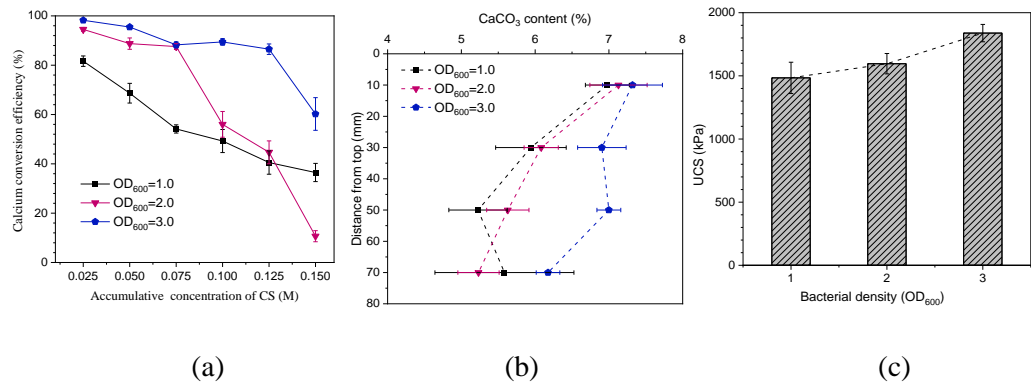

(a)      (b)      (c)

Figure 1 Effect of bacterial densities OD600 of 1.0, 2.0, and 3.0 on (a) chemical transformation efficiency, (b) CaCO3 content and (c) strength of MICP-treated soil

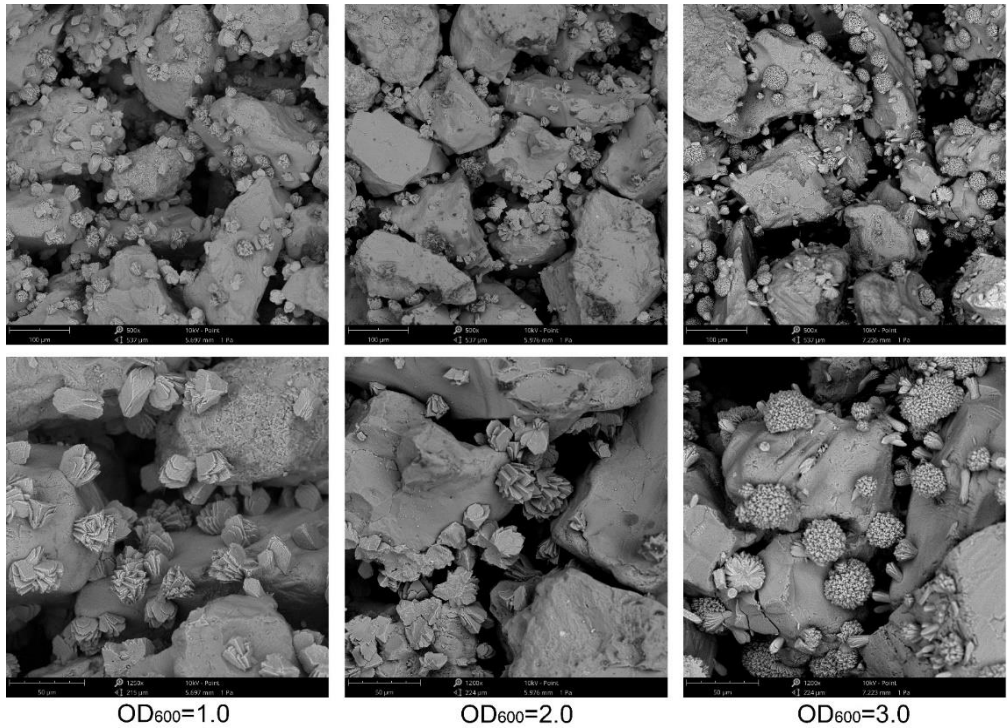

Figure 2 SEM images CaCO3 crystal properties via MICP treated by bacterial densities OD600 of 1.0, 2.0, and 3.0

Different bacterial retention times have been utilized in MICP research to examine the MICP performance, ranging from 2 hours to 24 hours. Figure 3 shows that the chemical conversion efficiency, $CaCO_3$ content, and UCS of samples treated with MICP are similar across retention times ranging from 2 to 24 hours. Therefore, a 2-hour retention time is sufficient to achieve a relatively high level of MICP efficiency for enhancing soil strength and can reduce the overall time required for MICP procedures compared to a 24-hour retention time. However, any retention time between 2 and 24 hours can be selected to align with the injection intervals between cementation solutions for engineering operational protocols.

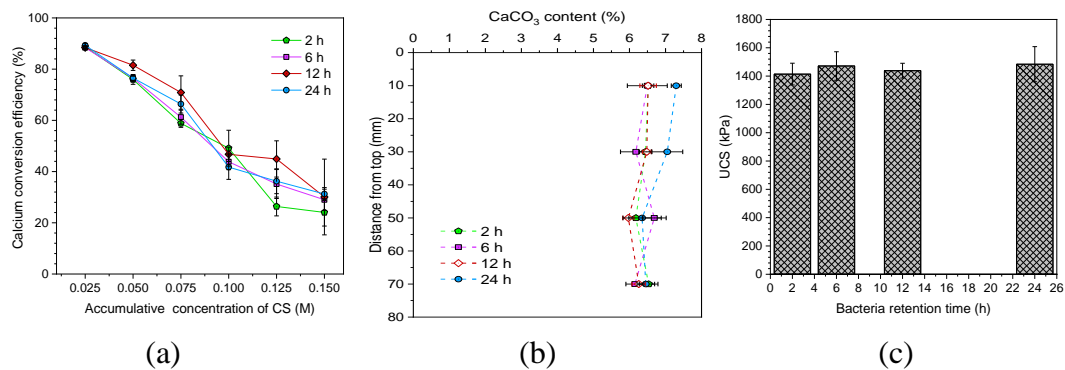

Figure 3 Effect of bacterial retention time (2, 6, 12 and 24 hours) on (a) chemical transformation efficiency, (b) CaCO3 content and (c) strength of MICP-treated soil

The effects of bacterial species on MICP were evaluated by comparing *S. pasteurii* and *S. aquamarina*. During the treatment process, the chemical transformation efficiency achieved with *S. aquamarina* was approximately 20%, significantly lower than the 40% to 80% range observed for *S. pasteurii* (see Figure 4). This difference is largely attributed to the initial activity levels of the bacteria prior to MICP treatment, with *S. aquamarina* exhibiting an activity of roughly 8.9±0.5 mM/h/OD compared to *S. pasteurii*'s significantly higher activity of about 39.4±0.5 mM/h/OD. The initial difference in chemical transformation efficiency between the two strains corresponds closely with the difference in their bacterial activity levels at the onset of MICP treatment.

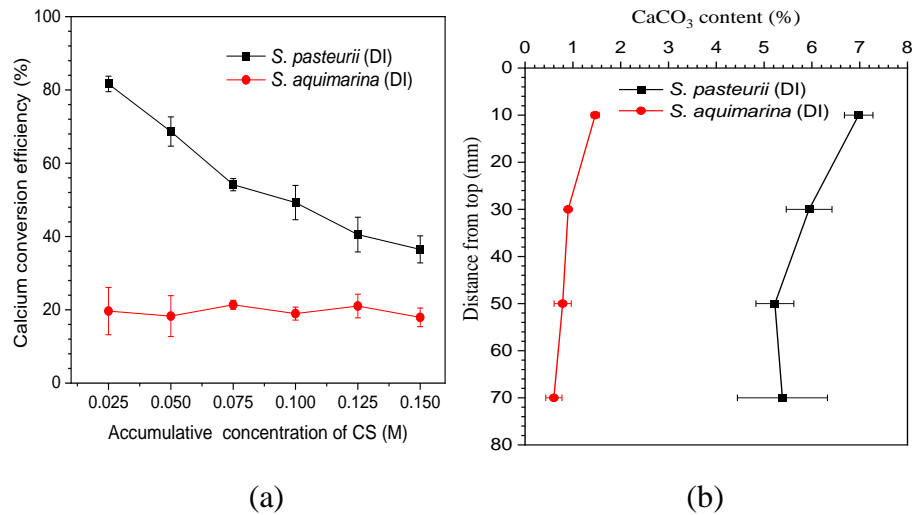

(a)  (b)

Figure 4 Effect of bacterial strain (S.pasteurii and S. aquimarina) on (a) chemical transformation efficiency and (b) CaCO$_3$ content of MICP-treated soil

**Cementation solution concentration**

The concentration of cementation solution for MICP has been reported to be between 0.1 M and 1.5 M. To achieve the same amount of CaCO$_3$ given that the chemical transformation efficiency is the same, the lower the concentration of cementation solution is, the more injection cycles are needed. However, the concentration of cementation solution affects the chemical transformation efficiency, and therefore an optimized concentration with moderate concentration and highest transform efficiency is need for MICP. Compared to relatively lower concentration of cementations solution such as 0.25 M and 0.5 M, higher concentrations such as 1.0 M and 1.5 M result in very low chemical transform efficiency (about 10-20%) (Figure 5a), and therefore much lower CaCO$_3$ content (about 0.2-1%) (Figure 5b) while the resulted UCS strength of the soil specimens was nearly zero (Figure 5c). The UCS values of soil treated by MICP with 0.5 M cementation solution is almost twice as that of 0.25 M (Figure 5c). The precipitated CaCO$_3$ crystals at 0.25M and 0.5 M cementation solution are quite similar, and 0.5 M produced slightly larger CaCO$_3$ crystals on average (Figure 6).

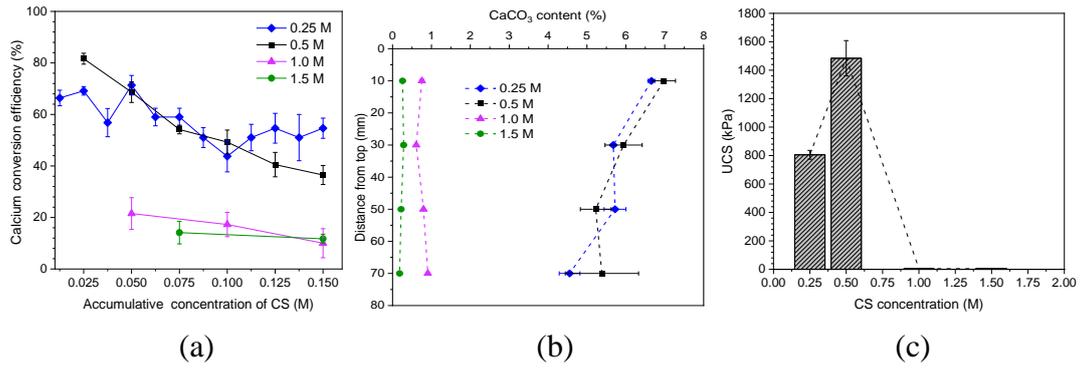

(a)            (b)            (c)

Figure 5 Effect of cementation solution concentration (2, 6, 12 and 24 hours) on (a) chemical transformation efficiency, $CaCO_3$ content and (c) strength of MICP-treated soil

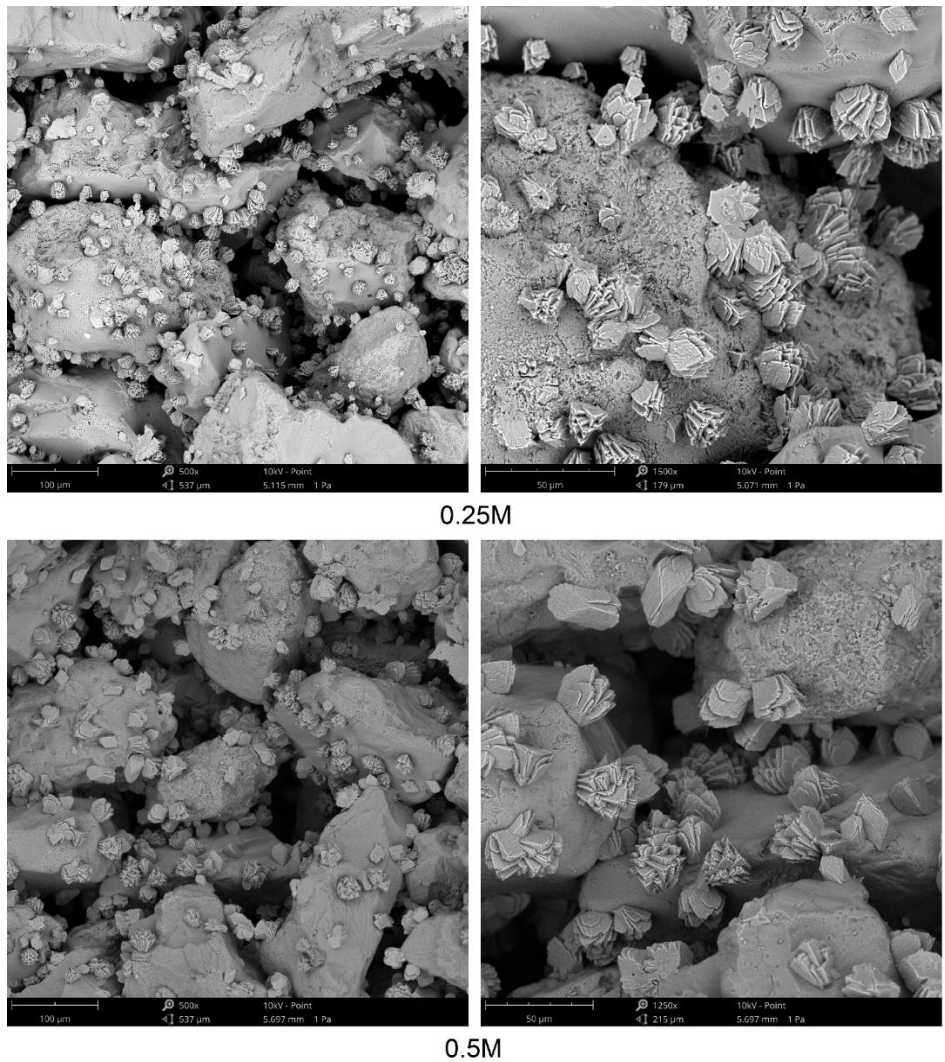

Figure 6 SEM images $CaCO_3$ crystal properties via MICP treated by cementation solution concentration (a) 0.25 M and (b) 0.5 M

**Salinity effects on bacterial strains**

When applying MICP in marine environments, it is important to consider the effect of salinity on the bacterial strain and the resulting MICP performance. To investigate this, two bacterial strains (*S. pasteurii* and *S. aquamarina*) were tested in both distilled water and seawater environments. Results show that the specimens generated with *S. aquamarina* have lower cementation levels (around 1% on average) compared to the specimens generated with *S. pasteurii* (6% on average) (Figure 7a), and the UCS values of the specimens treated with *S. aquamarina* regardless of exposed to seawater or not are almost zero (Figure 7b). The UCS values of the specimens treated with *S. pasteurii* exposed to seawater are slightly lower (1300 kPa) than those of the DI specimens (1500 kPa), but this difference is small, indicating the success of the method in marine environments and demonstrating the applicability of MICP (Figure 7b). The most significant difference observed in a saline environment is the type of crystals formed, as seen in the SEM images in Figure 8. While carbonate crystals in DI water have rhomboidal shapes as expected, in seawater they exhibit a mushroom-like shape. These findings, combined with the UCS results, suggest that the latter carbonate crystals are less efficient in terms of bridging particles and providing strength enhancement. Different carbonate crystal shapes under saline conditions were observed in other seawater-based biocementation studies in which these were characterised as irregular spherical particles or having an spherical shape (Yu and Rong, 2022; Cheng et al., 2014). Specifically, in the study by Yu and Rong (2022), EDS analysis was conducted showing that the cementation products are carbonates confirming the carbon being present in the crystals. Cheng et al. (2014), who utilised a very similar chemical recipe to generate artificial seawater as the one used in this work, performed EDX analysis confirming the presence of calcium being the second element present after silica (sand grains), and a very low amount of chlorine. It is recommended that future research includes conducting EDS (Energy Dispersive Spectroscopy) or XRD (X-ray Diffraction) analyses to investigate the composition of crystals formed, particularly when seawater or other salts are introduced into the MICP reaction system.

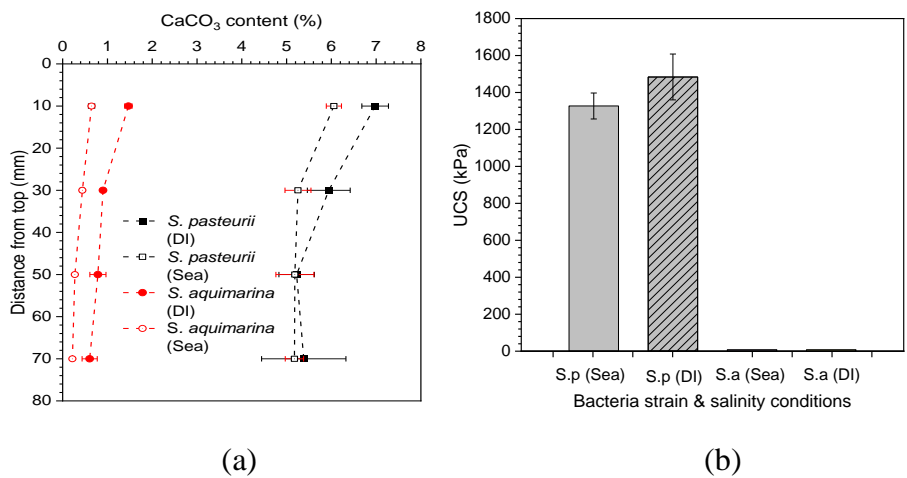

(a) (b)

Figure 7 Effects of bacteria and salts on (a) chemical transformation efficiency and (b) strength of MICP-treated soil

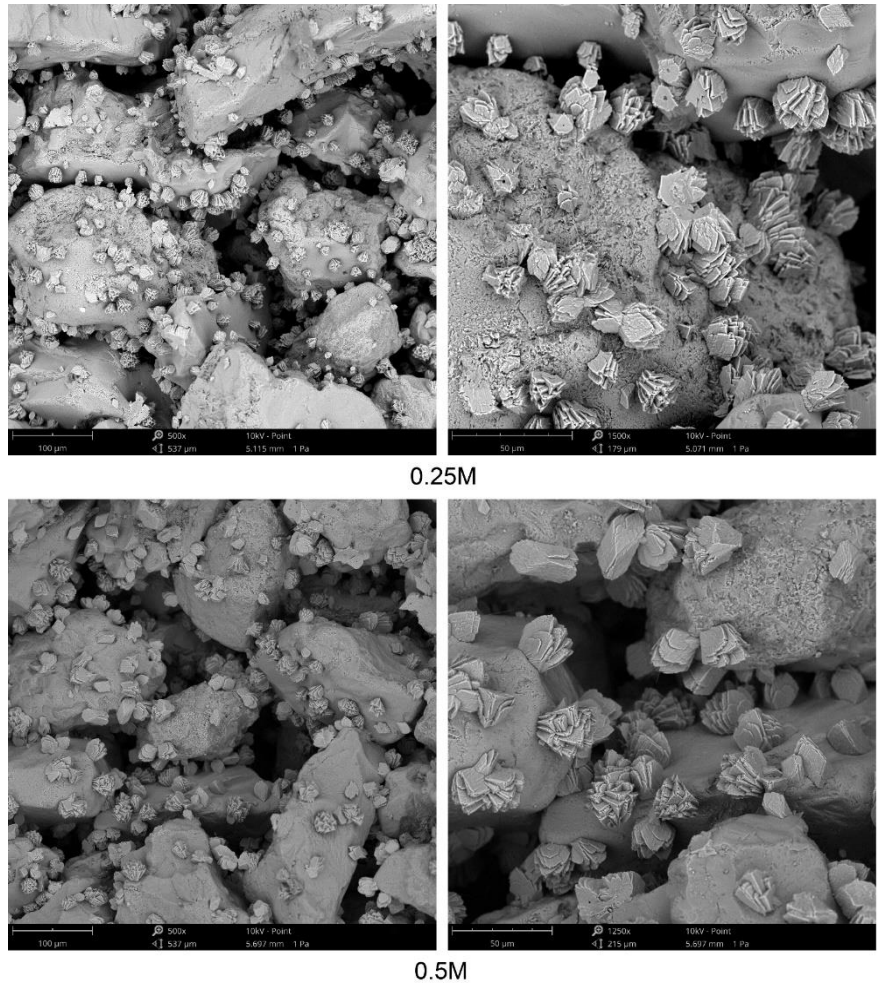

Figure 8 SEM images CaCO3 crystal properties via MICP treated by cementation solution made from (a) DI water and (b) seawater

**Environmental temperature, salinity and oxygen level**

The impact of temperature on bio-treatment and carbonate precipitation has been extensively studied under liquid conditions. In this study, experiments were conducted with *S. pasteurii* at temperatures of 4, 10, and 20°C, using DI water, at 10 and 20°C, with saline water, and at 20°C, with saline water both under aerobic and anaerobic conditions. 4°C, resulted in lowest calcium conversion efficiency, cementation levels and UCS (as seen in Figure 9a, b and c). At 10℃, the specimens treated with DI water had higher cementation levels (about 4.5% on average) compared to those treated with saline water (about 3.5% on average) as shown in Figure 9b. The optimal temperature for the *S. pasteurii* strain was found to be 20°C,, as both the cementation level and UCS were higher compared to specimens treated at 4 and 10°C,. The gap between the UCS values for seawater and DI water at 20°C, was much smaller under this temperature (about 1400 and 1500 kPa, respectively), as seen in Figure 9c. The SEM images presented in Figure 10 demonstrate that seawater facilitates the formation of small crystals, while the influence of oxygen levels on the morphology of calcium carbonate is minimal.

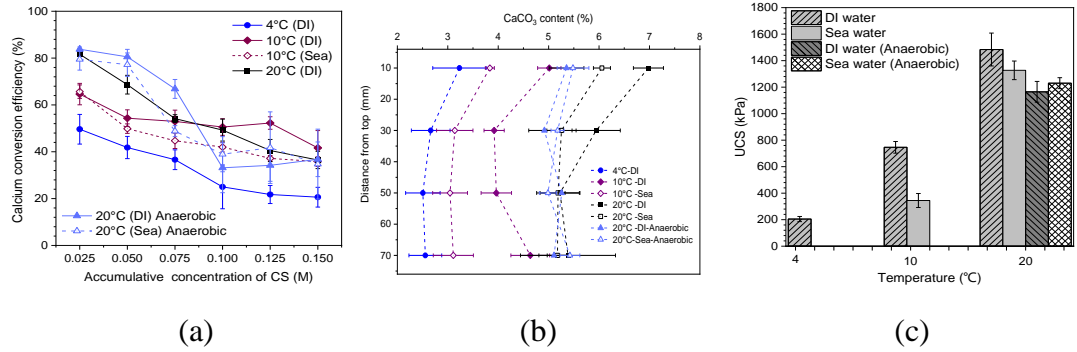

Figure 9 Effects of temperature, salinity and oxygen on (a) chemical transformation efficiency, (b) CaCO$_3$ content and (c) strength of MICP-treated soil

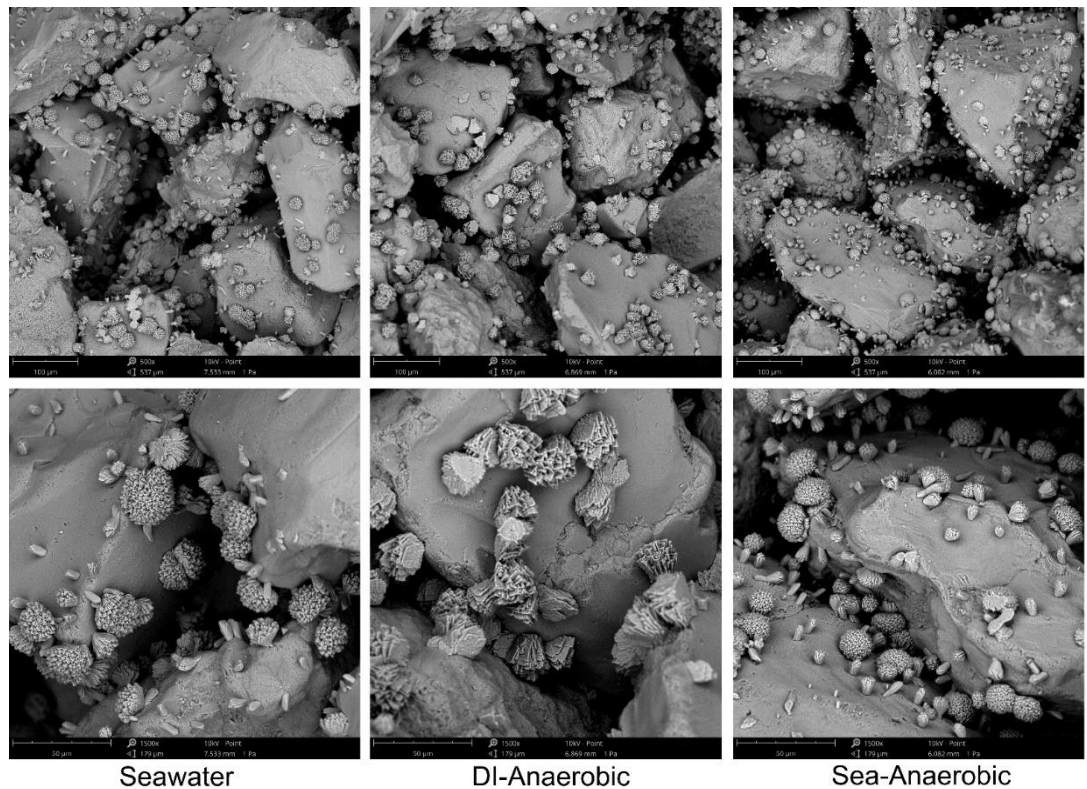

Figure 10 SEM images CaCO3 crystal properties via MICP treated under saline, aerobic or anaerobic conditions

## Number of bacterial injections

In the work by Wang et al. (2022), it is suggested that utilizing multiple bacterial solution (BS) injections is an effective strategy to achieve higher strength at high temperatures using the specific bacterial strain. This hypothesis is examined herein for low temperature condition (Figure 11) and for low activity bacterial strain (Figure 12). As depicted in Figure 11, the amount of cementation for a single BS injection is lower (around 2.5% on average) and less uniform, while for the other two cases, it is higher (around 4%) and the specimens are relatively uniform. The UCS values for BS numbers of 3 and 6 are 550 and 500 kPa, respectively, indicating that there is no significant difference between the two. However, the difference between a single BS injection and multiple BS injections is significant, with the former resulting in a UCS value of only 200 kPa on average.

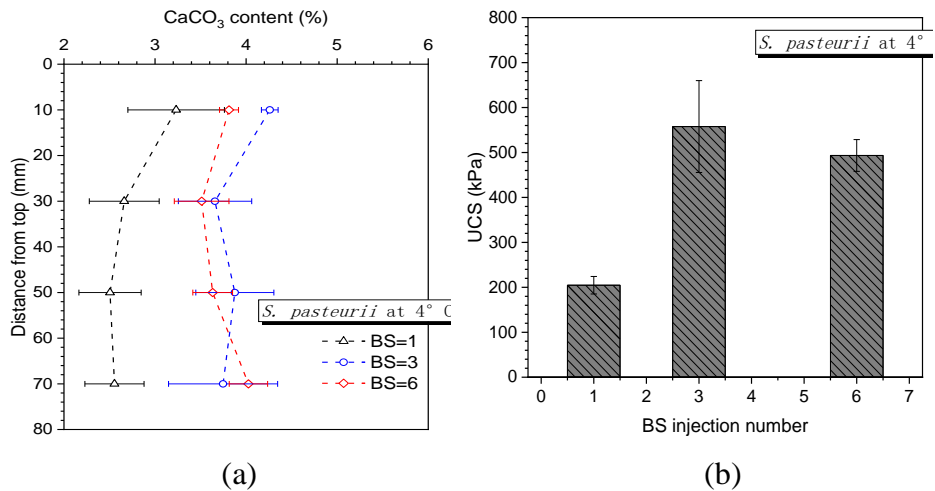

(a)      (b)

Figure 11 Effect of bacterial suspension injection number for specimens treated with the S.pasteurii on (a) $CaCO_3$ content and (b) strength of MICP-treated soil

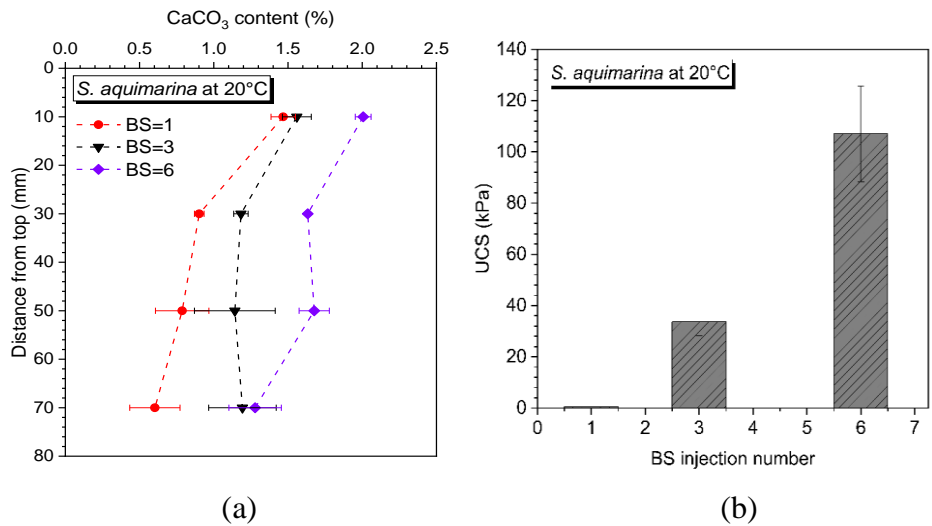

(a)      (b)

Figure 12 Effect of bacterial suspension injection number for specimens treated with the S. aquimarina on (a) $CaCO_3$ content and (b) strength of MICP-treated soil

The effects of increasing bacterial injection number for increasing MICP efficiency by bacterial strain *S. aquimarina* are significant, as with the BS injection number increase from 1 to 6 average CaCO$_3$ content increases (Figure 12a) and UCS increases (Figure 12b). The UCS values increased from almost zero when only once of bacterial injection was conducted to about 350 kPa when 6 times of bacterial suspension is injected.

**pH controlled (various pH levels) one phase injection**

For one-phase injection procedure, the effects of different pH levels, 6.0, 7.5, and 9.0, on the chemical conversion efficiency, uniformity, and strength of specimens were studied. Results showed that for all three pH levels, chemical transformation efficiency increased with the injection of cementation solution, with about 80% efficiency achieved at 1.5 M of CS, by injecting 6 times of mixtures of bacterial suspension and cementation solution (Figure 13a). This is different to the other treatment protocol where when the bacterial suspension and cementation solution were injected sequentially the chemical transformation efficiency decreased with the increase of cementation solution injection number (Figure 1a, Figure 5a, and Figure 9a). In addition, pH had a significant impact on specimen uniformity, with specimens treated at pH levels of 6.0 and 7.5 showing relatively similar cementation levels across the four tested points, while specimens treated at a pH of 9.0 exhibited great variation in cementation levels (Figure 13b). The strength of the specimens was greatly enhanced at a pH of 7.5, with an average value of 800 kPa, followed by the specimen treated at a pH of 6.0, with a UCS value of 650 kPa. The specimen treated at a pH of 9.0 had a very low strength, just below 200 kPa (Figure 13c). Although the specimen treated at a pH of 6.0 was more uniform than the one treated at 7.5, the latter condition generated larger carbonate crystals, which provided better grain-to-grain bridges (Figure 14) and thus higher strength (Figure 14c). At a pH of 6.0, the carbonate crystals appeared elongated and less rounded (Figure 14). At a pH of 9.0, the carbonate crystals remained round and not big enough to bond effectively bond soil particles (Figure 14), which are less efficient in increasing the friction angle compared to the elongated shape, and due to the not sufficient bonding and overall low CaCO$_3$ content, the strength was lower (Figure 13c).

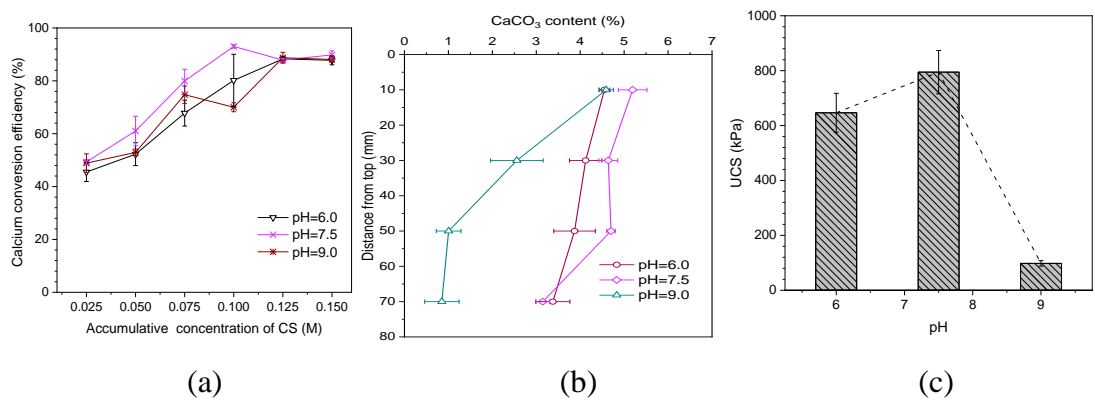

(a) (b) (c)

Figure 13 Effect of *pH* (6.0, 7.5 and 9.0) on (a) chemical transformation efficiency, (b) CaCO$_3$ content and (c) strength of MICP-treated soil

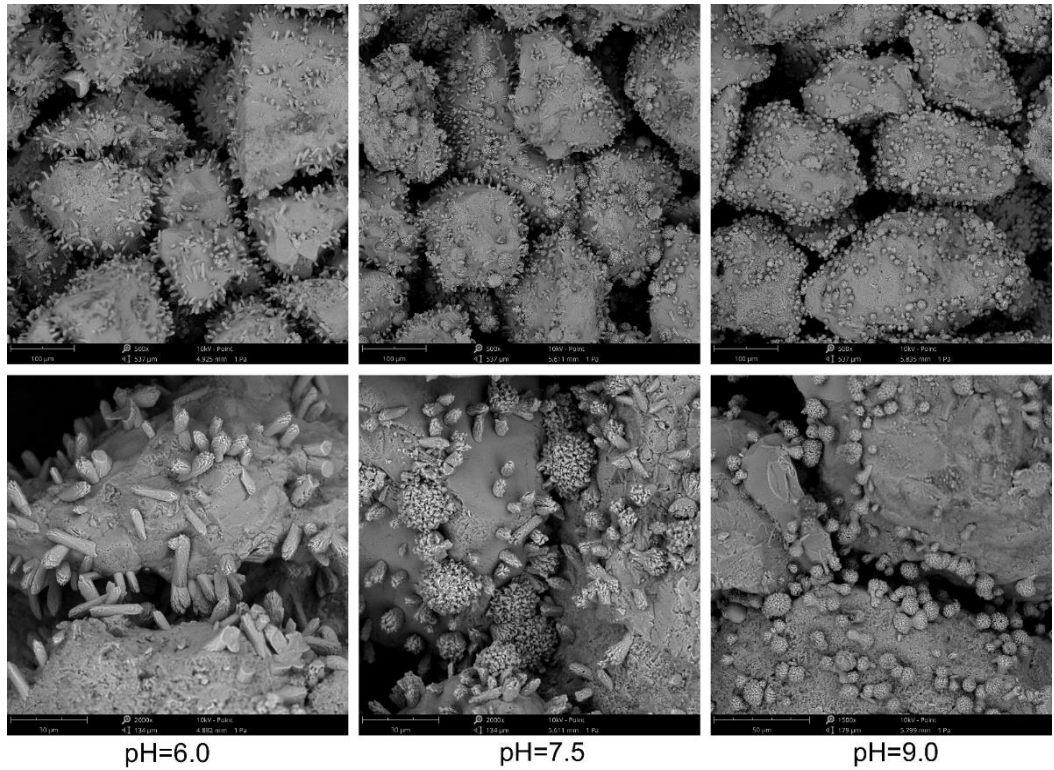

Figure 14 SEM images CaCO3 crystal properties via MICP treated by pH 6.0, 7.5 and 9.0

## 4. Discussion

**Analysis of factor significance**

Table 4 and Figure 15 summarize the Calcium Carbonate Content (CCC) and UCS of specimens generated under various bio-chemical protocols and influenced by diverse environmental conditions, facilitating a comparison of factors affecting MICP. Table 4 also summarizes the chemical transformation data. Additionally, Figure 15, along with the SEM images presented in the Results section, indicates the relationship between UCS and CCC (see section 4.2) influencing the macro-scale strength response.

Table 4 Summary of factors considered and their effects on CCE, CCC and UCS

| Factors | | CCE (%) | | CCC (%) | | UCS (kPa) | |
|---|---|---|---|---|---|---|---|
| | | Average | Range | Average | Range | Average | Range |
| Bio-chemical factors | | | | | | | |
| OD$_{600}$ | 1 | 55.12 | 31.23 | 5.93 | 0.92 | 1484 | 354.5 |
| | 2 | 63.7 | | 6.02 | | 1596 | |
| | 3 | 86.35 | | 6.85 | | 1838.5 | |
| Bacterial retention time (h) | 2 | 53.84 | 6.58 | 6.42 | 0.48 | 1414 | 70 |
| | 6 | 55.67 | | 6.38 | | 1471.33 | |
| | 12 | 60.42 | | 6.31 | | 1438.33 | |
| | 24 | 56.9 | | 6.79 | | 1484 | |
| Bacterial strain | S. p | 55.11 | 35.58 | 5.88 | 4.94 | - | - |
| | S. a | 19.54 | | 0.94 | | - | |
| Concentration of cementation solution (M) | 0.25 | 58.76 | 45.84 | 5.65 | 5.65 | 805.33 | 1479 |
| | 0.5 | 55.11 | | 5.88 | | 1484 | |
| | 1 | 16.28 | | 0.77 | | 5 | |
| | 1.5 | 12.92 | | 0.24 | | 5 | |
| Environmental factors | | | | | | | |
| Bacteria (S.p/S.a)-salinity(Yes/No, Y/N) | S.p-N | - | - | 5.88 | 5.49 | 1484 | 1476 |
| | S.p-Y | - | | 5.42 | | 1327 | |
| | S.a-N | - | | 0.94 | | 8 | |
| | S.a-Y | - | | 0.4 | | 8 | |
| Temperature (°C)−salinity(Yes/No, Y/N)- oxygen (Aerobic/Anaerobic, Ae/An) | 4-N-Ae | 32.56 | 23.32 | 2.74 | 3.14 | 204.67 | 1279.33 |
| | 10-N-Ae | 52.73 | | 4.38 | | 746 | |
| | 10-Y-Ae | 45.86 | | 3.28 | | 344.67 | |
| | 20-Y-Ae | 55.12 | | 5.42 | | 1327 | |
| | 20-N-An | 55.89 | | 5.16 | | 1165 | |
| | 20-Y-An | 53.48 | | 5.27 | | 1230 | |
| Protocol factors | | | | | | | |
| Bacterial injection number at 4°C | 1 | - | - | 2.74 | 1.15 | 204.67 | 353 |
| | 3 | - | | 3.89 | | 557.67 | |
| | 6 | - | | 3.75 | | 493.33 | |
| Bacterial injection number for *S. a* | 1 | - | - | 0.94 | 0.71 | 0.73 | 106.37 |
| | 3 | - | | 1.27 | | 33.91 | |
| | 6 | - | | 1.65 | | 107.1 | |
| One-phase injection pH | pH 6.0 | 70.3 | 6.53 | 3.98 | 2.17 | 646.5 | 696.75 |
| | pH 7.5 | 76.83 | | 4.42 | | 794.75 | |
| | pH 9.0 | 70.57 | | 2.25 | | 98 | |

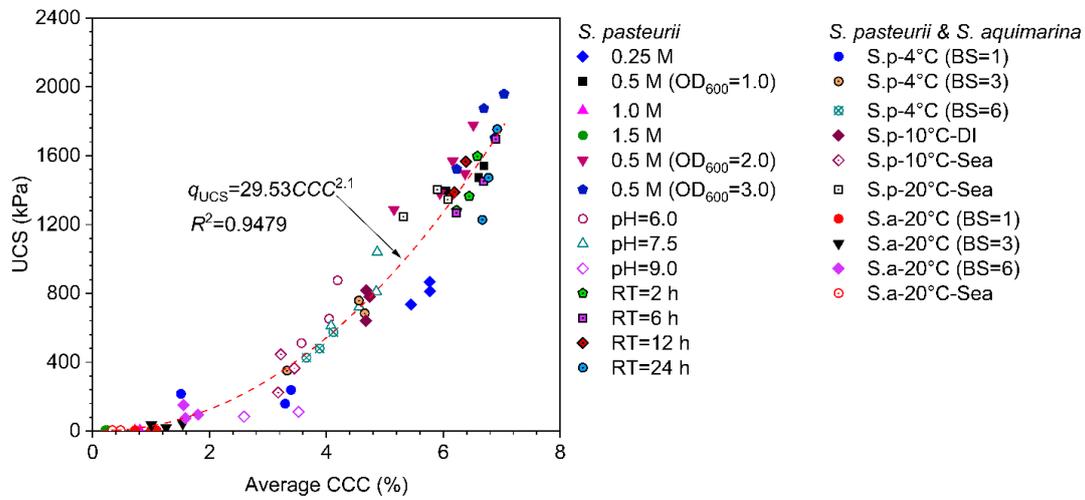

Figure 15 Correlation between average calcium carbonate content and UCS

The significant variation in CCC and UCS highlights different outcomes even with the same amount of chemical injection, as detailed in Table 4, where CCC varies from zero to approximately 7.5% and UCS from zero to around 2 MPa. These deviations are attributed to the different bio-chemical factors selected. In particular, the concentration of the cementation solution and the selection of bacterial species stand out as key factors affecting MICP performance. Among bacterial strains, *S. pasteurii* is noted for its adaptability in various environments, whereas *S. aquimarina* shows limited effectiveness for MICP, regardless of the environmental setting, despite its marine origins. A cementation solution concentration of 0.5 M is identified as most efficient among the tested concentrations (0.25, 0.5, 1.0, and 1.5 M), with 0.25 M following, while concentrations of 1.0 M or 1.5 M are less suitable. In addition to bacterial species and cementation solution concentration, other bio-chemical factors such as bacterial density and retention times were investigated. While optical densities ranging from 1.0 to 3.0 were proved effective for MICP in soil column studies, exploring higher bacterial densities becomes essential for larger-scale soil treatments to avoid aggregation. Bacterial densities ranging from 1.0 to 3.0 were found suitable for soil column tests. Retention times of bacteria between 2 to 24 hours showed comparable efficiency in MICP and soil strength enhancement, offering flexibility for real-world application designs.

Among the different environmental factors studied (temperature, oxygen condition, and salinity), temperature emerges as the most influential factor affecting MICP performance. Within the temperature range of 4°C, 10°C and 20°C, there is an almost linear enhancement in both the average $CaCO_3$ content throughout the sample lengths and the UCS values (Figure 9 b and c). The impact of salt on MICP performance is less pronounced than that of temperature and also varies depending on the temperature. At 20°C, the addition of sea salt to the cementation solution has a minor detrimental effect on MICP efficiency, which can still be effectively managed. In contrast, at 10°C, the presence of sea salt in the cementation solution substantially reduces MICP efficiency by half, as illustrated in Figure 9c. The absence of oxygen in the DI (Deionized Water) system results in a more pronounced decrease of 21.4% in UCS compared to when oxygen is available, whereas in the seawater system, the decrease in UCS is 7% when oxygen is unavailable (Figure 9c).

Recently, one-phase MICP treatment has emerged as a promising alternative MICP protocol due to its advantage of simplifying treatment protocols compared to two-phase injections (Cheng et al. 2019; Cui et al. 2022; Lai et al. 2022). However, one-phase treatments with pH values of 6.0 or 7.5 yielded CCC levels of 3.5-5% and UCS values of 500-1000 kPa. Nonetheless, these values were found to be lower than those achieved through two-phase injections, where factors such as cementation solution concentration and temperature were held constant. Conversely, a pH of 9.0 led to the poorest MICP performance due to excessive precipitation occurring before injection into the soils. Hence, in MICP applications, achieving a balance between the complexity of MICP treatment and its efficiency is crucial, especially for large-scale *in situ* treatments. Furthermore, in order to enhance MICP performance at high temperatures, recent studies have proposed modifications to the two-phase injection method by increasing the number of bacterial injections compared to traditional single injections (Wang et al. 2022, 2023). Based on this investigation, the present study indicates that although increasing the injection frequency in cases involving 4°C and *S. aquimarina* can enhance MICP performance to some degree (as depicted in Figure 12), the efficacy remains significantly inferior compared to cases involving 20°C or *S. pasteurii* (Figures1 and 15). Thus, these findings suggest that the influence of bio-chemical factors, such as selecting the appropriate bacterial strain, and environmental factors such as 20°C, have a more significant impact compared to optimizing protocols by increasing bacterial injections using a strain with low activity, or conducting the MICP process under low temperature conditions such as 4°C.

**Effects of $CaCO_3$ content and micro-scale properties on enhancing soil strength**

In the experiment, it was observed that an increase in $CaCO_3$ content, generated through MICP, generally leads to higher UCS of the soil (Figure 15). However, when the $CaCO_3$ content remains constant, variations in UCS were noted among samples influenced by different biochemical or environmental factors. Alongside the results of SEM images, this suggests that both the $CaCO_3$ content and its microscale properties play a crucial role in influencing the strength of MICP-treated soils.

Earlier research has proposed that larger $CaCO_3$ crystals, which form stronger bonds with soil particles, are more efficient in enhancing the strength of MICP (Cheng et al. 2017). This argument was validated in the present study. In the three cases treated with one-phase different pH levels, although the specimen treated at a pH of 6.0 exhibited greater uniformity compared to the one treated at 7.5, the latter condition resulted in the formation of larger carbonate crystals (Figure 14), which in turn facilitated better grain-to-grain bridges (Figure 15). Furthermore, at a pH of 9.0, the strength of MICP-treated samples fell towards the lower end of the fitting line, indicating insufficient bonding, as shown in the SEM images presented in Figure 14. In the two cases where the cementation solution varied (Figure 6), the use of 0.5 M cementation solution led to the formation of larger crystals compared to the 0.25 M case (Figure 6). Consequently, the crystals produced by the 0.5 M solution more effectively bonded soil particles and enhanced strength (as indicated by the dots clustered around the fitting line in Figure 15). Conversely, the 0.25 M solution exhibited less effective bonding of soil particles and consequently less strength enhancement, as evidenced by the scattering of dots below the fitting line in Figure 15.

Previous studies have indicated that during MICP procedures, the morphology of the precipitated $CaCO_3$ can vary depending on bacterial density (Wang et al. 2019b and 2021). When the bacterial density is relatively high, such as 1.0 and 3.0, it may transition from a morphology resembling Amorphous Calcium Carbonate (ACC) to vaterite, and then to calcite (Wang et al., 2021). Post-MICP treatment, the predominant observed crystal types are typically vaterite and calcite (van Paassen, 2010). These crystal types exhibit distinct shapes; for instance, vaterite tends to be round-shaped, either hollow or dense (van Paassen, 2010), while calcite crystals are rhombohedral (van Paassen, 2010), which can continue growing into rounded form (Wang et al. 2021) as well. Notably, hollow crystals are detected when the concentration of cementation solution is 1.0 M, suggesting that these hollow crystals may not adequately bond with soil particles (van Paassen, 2010).

**Effects of bio-chemical factors on bacterial activity and the rate of MICP chemical reactions**

Factors such as temperature can influence bacterial activity, thereby indirectly affecting MICP efficiency, or directly impacting the rate of MICP chemical reactions. Wang et al. (2022, 2023) conducted both microfluidic chip experiments and soil column experiments to investigate the influence of environmental temperature on MICP. Their findings indicate that temperature variations yield differing effects. At elevated temperatures, such as 50°C, bacterial density decreases significantly over time. Consequently, despite potentially higher precipitation rates at higher temperatures, the overall bio-chemical reaction rate is notably lower due to reduced bacterial activity compared to room temperature. Conversely, at lower temperatures, such as 4°C, bacterial activity remains relatively high throughout the experiment period. However, the total bio-chemical reaction rate is still lower than at room temperature, as lower temperatures directly hinder precipitation chemical reactions, despite maintaining a high ureolysis rate. Therefore, at 50°C, MICP efficiency can be enhanced by additional bacterial injections, ensuring a relatively high ureolysis rate throughout the precipitation period (Wang et al., 2022, 2023). Conversely, at lower temperatures, such as 4°C as demonstrated in the current study, increasing bacterial injections from 1 to 3 leads to a rise in UCS from 200 kPa to 550 kPa. However, further increasing the injection number from 3 to 6 does not result in a higher UCS; instead, it remains similar at 500 kPa.

Moreover, the present study also demonstrates the significance of increasing bacterial injection numbers for enhancing MICP efficiency in cases with low-activity bacteria, such as *S. aquimarina*. As the number of bacterial injections (BS) increases from 1 to 6, both the average $CaCO_3$ content (Figure 12a) and UCS values (Figure 12b) show notable improvements. Initially, UCS values are almost zero with only one bacterial injection, but they increase to approximately 100 kPa with three injections of bacterial suspension and further to around 350 kPa with six injections. This trend resembles the findings in high-temperature cases, where augmenting bacterial activity through additional bacterial injections significantly enhances MICP efficiency.

Therefore, the efficacy of multiple injections in effectively enhancing MICP performance relies on specific conditions. In instances where bacterial activity is low, as observed in cases such as 50°C and involving *S. aquimarina*, multiple injections can indeed enhance MICP performance. However, in situations where low MICP performance does not result from reduced bacterial activity but rather, results from a low chemical reaction rate, as seen in scenarios at 4°C, multiple injections are proven ineffective. Moreover, understanding the key mechanisms influencing MICP biological and chemical rates can aid in modifying MICP treatment procedures to achieve optimum performance.

In addition to temperature, which directly and indirectly influences MICP efficiency, the concentration of the cementation solution may also affect the chemical efficiency. In theory, a higher concentration of $Ca^{2+}$ and $CO_3^{2-}$ ions should result in a higher precipitation rate. However, the current study reveals that when the concentration of the cementation solution is 1.0 M or 1.5 M, MICP efficiency is significantly lower (only 20%) compared to when the concentration is 0.5 M (around 80-40%) (Figure 5a). Studies have shown that higher $Ca^{2+}$ concentrations reduce bacterial activity, consequently reducing the production rate of $CO_3^{2-}$ (van Paassen 2010). Thus, even with a high $Ca^{2+}$ concentration, if there is not a corresponding high concentration of $CO_3^{2-}$ produced, the precipitation rate remains low. Additionally, studies have demonstrated that elevated $Ca^{2+}$ levels can lead to bacterial aggregation (El Mountassir et al., 2014, Wang 2018), which may contribute to both a lower overall ureolysis rate and clogging (as discussed in Section 4.4), consequently reducing MICP efficiency.

**Effects of MICP protocols and environmental factors on permeability reduction**

Figure 16 presents the measured flow rate calculated from outflow in each injection event against $CaCO_3$ content. Generally, the flow rate decreased from 0.45 cm$^3$/s to approximately 0.1 cm$^3$/s. However, the reduction in flow rate with increasing $CaCO_3$ content exhibited significant variability among the samples. Particularly, samples treated with 1.0 M and 1.5 M cementation solutions, as well as the one-phase pH 9 experiment, displayed a sharp reduction in flow rate, indicating local clogging, as suggested by Konstantinou et al. (2021b). The initial flow rate showed variation across different samples, ranging from approximately 2.5 to 4.5, attributable to the diversity in soil characteristics. For future investigations in which the focus is on permeability or hydraulic conductivity, the utilization of micro-CT technique is recommended, as it enables direct observation of the microstructure of the soil skeleton. This approach can offer valuable insights into understanding the complex and spatial permeability changes caused by MICP treatment.

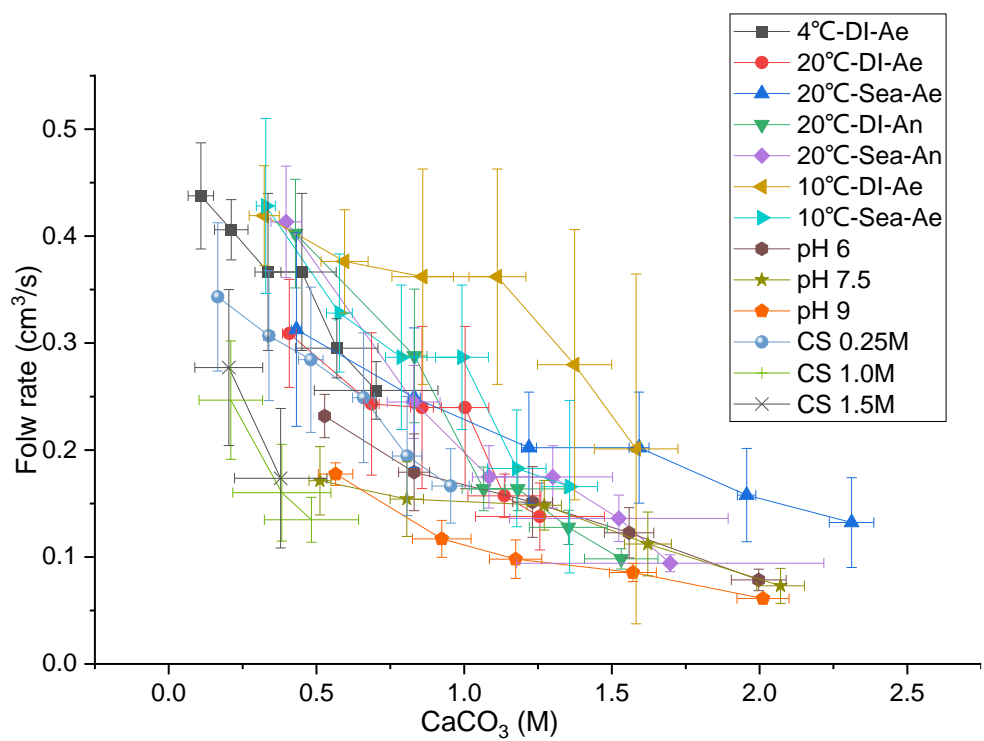

Figure 16 Correlation between flow rate and calculated CaCO3 content

**Optimal MICP protocols**

Upon evaluation of the various bio-chemical and environmental factors in this study, the most favorable MICP performance was observed in samples treated with *S. pasteurii* bacteria, utilizing a 0.5 M cementation solution, and maintaining a temperature of 20℃. Previous studies have also demonstrated that a temperature of 35℃ is favored for MICP (Wang et al., 2022). While a 0.25 M cementation solution is nearly as effective as a 0.5 M solution for MICP, since it requires doubling the injection amount, the 0.5 M solution is the most recommended. However, optical densities of 1.0-3.0 all yield good results, with 3.0 being the highest. Nonetheless, larger scale experiments should be conducted to confirm the optimal bacterial density, as higher bacterial densities might result in non-homogeneous distribution of bacteria, and even bacterial clogging, which could reduce overall performance. Environmental salinity and oxygen do not significantly affect MICP performance at the column scale presented in the current study; however, larger scale experiments need to be conducted to study their effects spatiotemporally. Even in marine environments, *S. pasteurii* remains the most efficient bacterial strain compared to *S. aquimarina* , which are abundant in marine environments.

## 5. Conclusions

This study introduces a comprehensive program designed to evaluate the impacts of various bio-chemical and environmental factors on MICP performance. These factors include bacterial strains, bacterial density, retention times, chemical solution concentrations, and environmental parameters such as temperature, oxygen levels, seawater salinity, and pH. The key findings are summarized below.

Among the investigated bio-chemical factors pertinent to Microbially Induced Calcium Carbonate Precipitation (MICP)—namely, bacterial strains, density, retention times, and chemical solution concentrations—it is evident that bacterial strain and chemical solution concentrations have significant effects on MICP performance. Moreover, bacterial density also plays a crucial role, particularly for less active bacterial strains, where increased injection of bacteria enhances MICP efficacy. Conversely, retention times ranging from 2 to 24 hours have marginal effects. *S. pasteurii* is the preferred MICP strain, exhibiting optimal performance at $OD_{600}$ 1.0, with marginal improvements at optical densities of 2.0 and 3.0. However, other ureolytic bacteria such as *S. aquamarine* display lower efficacy, even with multiple injections. Chemical concentrations of 0.25 M and 0.5 M result in higher chemical transformation efficiencies and $CaCO_3$ contents, whereas concentrations of 1.0 M and 1.5 M result in chemical transformation efficiencies lower than 20%.

Among the environmental factors affecting MICP—temperature, oxygen levels, and seawater salinity—temperature within the range of 4 to 20°C demonstrates a higher impact on the MICP performance compared to salinity and oxygen. UCS values exhibit a linear increase with temperatures within this range. Following Wang et al. (2023), temperatures ranging from 20 to 35°C prove to be suitable for MICP processes. While single-phase injection is less complex than two-phase injection, the optimal performance observed in the single-phase method, particularly when pH is adjusted to 6 or 7.5, yields MICP-treated UCS values only approximately half as effective as those achieved through the two-phase method under identical conditions.

The change in flow rate due to increasing $CaCO_3$ content had noticeable variation in the samples. Notably, samples treated with 1.0 M and 1.5 M cementation solutions, along with the pH 9 one-phase experiment, experienced a significant flow rate decrease, suggesting potential local clogging. The initial flow rate varied among samples, ranging from around 2.5 to 4.5, reflecting the diversity in soil characteristics.

Utilizing soil column experiments and UCS measurements represents an established method for studying MICP. This approach proves valuable in evaluating the performance of MICP-treated soils under varying biochemical and environmental conditions. Nevertheless, it is advisable to conduct larger-scale experiments in the future to investigate deeper into the effects of various factors. Understanding MICP performance across diverse factor combinations aids in designing effective MICP treatment protocols for different applications.